\documentclass[prd,nofootinbib,12pt]  {revtex4}
\usepackage{amsfonts,amsmath,amscd}
\usepackage{graphicx}

\begin{document}
\small
\title{%
Gauged AdS--Maxwell algebra and gravity}
\author{R. Durka}
\email{rdurka@ift.uni.wroc.pl}\affiliation{Institute for Theoretical
Physics, University of Wroc\l{}aw, Pl.\ Maksa Borna 9, Pl--50-204
Wroc\l{}aw, Poland}
\author{J. Kowalski-Glikman}
\email{jkowalskiglikman@ift.uni.wroc.pl}\affiliation{Institute for
Theoretical Physics, University of Wroc\l{}aw, Pl.\ Maksa Borna 9,
Pl--50-204 Wroc\l{}aw, Poland}
\author{M. Szcz\c{a}chor}
\email{misza@ift.uni.wroc.pl} \affiliation{Institute for Theoretical
Physics, University of Wroc\l{}aw, Pl.\ Maksa Borna 9, Pl--50-204
Wroc\l{}aw, Poland}

\date{\today}

\begin{abstract} We deform the anti-de Sitter algebra  by adding  additional generators $\mathcal{Z}_{ab}$,
forming in this way the negative cosmological constant counterpart
of the Maxwell algebra. We gauge this algebra and construct a
dynamical model with the help of a constrained the BF theory. It
turns out that the resulting theory is described by the
Einstein-Cartan action with Holst term, and the gauge fields
associated with the Maxwell generators $\mathcal{Z}_{ab}$ appear
only in topological terms that do not influence dynamical field
equations. We briefly comment on the extension of this construction,
which would lead to a nontrivial Maxwell fields dynamics.
\end{abstract}

\maketitle

The Maxwell algebra is a non-central extension of Poincar\'e algebra
obtained by replacing the commutator of translations $[P_a, P_b]=0$
with
\begin{equation}\label{1}
    [\mathcal{P}_a, \mathcal{P}_b] = i \mathcal{Z}_{ab}\, ,
\end{equation}
with $Z_{ab}=-Z_{ba}$ being six abelian generators commuting with
translations and forming  a tensor with respect to Lorentz
transformations
\begin{equation}\label{2}
    [\mathcal{M}_{ab},\mathcal{Z}_{cd}]=-i(\eta_{ac}\mathcal{Z}_{bd}+\eta_{bd}\mathcal{Z}_{ac}-\eta_{ad}\mathcal{Z}_{bc}-\eta_{bc}\mathcal{Z}_{ad})
\end{equation}
  Such generalization of Poincar\'e algebra arises
when one considers symmetries of systems evolving in flat Minkowski
space filled in by constant electromagnetic background
\cite{Bacry:1970ye}, \cite{Schrader:1972zd}. This kind of extension
of the Poincar\'e algebra is also of purely algebraic interest
because it circumvents a well known theorem that does not allow for
central extension of this algebra (see e.g., \cite{Galindo:1967}, \cite{Soroka:2011tc}, \cite{Gomis:2009dm}).

The Maxwell algebra attracted some attention recently because its
supersymmetrization leads to a new form of the supersymmetry $\mathcal N=1$,
$D=4$ algebra, containing the super-Poincar\'e algebra as its
subalgebra \cite{Bonanos:2009wy}. Even more interestingly it has
been argued in \cite{deAzcarraga:2010sw} that by making use of the
gauged Maxwell algebra one can understand it as a source of an
additional contribution to the cosmological term in Einstein
gravity. In this paper we would like reexamine this claim. To this
aim we present here an alternative construction of the action of
gravity based on the gauging of the AdS-Maxwell algebra employing
the concept of  a constrained BF theory.

It is well known that the action for gravity can be written in the
form of a constrained BF theory for the de Sitter or Anti de Sitter
algebra \cite{Starodubtsev:2003xq}, \cite{Smolin:2003qu},
\cite{Freidel:2005ak}, and \cite{Wise:2006sm}. Let us shortly review
this construction in the AdS case (the dS counterpart can be
constructed along the same lines.)

Take the Anti de Sitter algebra $\sf{so}(3,2)$
\begin{equation}\label{3}
    [\mathcal{M}_{IJ},\mathcal{M}_{KL}]=i(\eta_{IL}\mathcal{M}_{JK}+\eta_{JK}\mathcal{M}_{IL}-\eta_{IK}\mathcal{M}_{JL}-\eta_{JL}\mathcal{M}_{IK})\,
\end{equation}
with the metric tensor $\eta_{IJ}$, $I,J=0,\ldots,4$ having the
signature $(-,+,+,+,-)$. Consider the connection one form $A^{IJ}$
and the two form field $B^{IJ}$, both valued in this algebra, and
take the most general Lagrangian quadratic in the field $B$ and the
curvature two form $F^{IJ}$ of the connection $A^{IJ}$. The action
reads
\begin{equation}\label{4}
 16\pi \, S(A,B)= \int F^{IJ}\wedge B_{IJ} -\frac{\beta}{2} B^{IJ}\wedge B_{IJ} - \frac{\alpha}{4}\epsilon^{IJKL4} B_{IJ}\wedge B_{KL}
\end{equation}
with $\alpha$ and $\beta$ being dimensionless coupling constants.
The first two terms in this action are invariant under the action of
local $\sf{so}(3,2)$ gauge symmetries if $B^{IJ}$ transform under
these symmetries like curvature (see below). The third term,
however, is invariant only under the action of a subgroup of the
Anti de Sitter group which leaves $ \epsilon^{IJKL4}$
invariant\footnote{As it is well known the totally antisymmetric
symbol $ \epsilon^{IJKLM}$, defined by $ \epsilon^{01234}=1$, is an
invariant tensor of the algebra $\sf{so}(3,2)$.}, which is its
Lorentz subgroup with the algebra $\sf{so}(3,1)$. This term can be
thought of as a constraint, explicitly breaking the local
translational invariance and rendering the action only local-Lorentz
invariant.

It is a remarkable fact that the action (\ref{4}) is equivalent to
the action of Einstein-Cartan gravity (with negative cosmological
constant) appended by topological terms. To see this one decomposes
the connection $A^{IJ}$ into Lorentz connection and the tetrad field
\begin{equation}\label{5}
A^{ab}_\mu=\omega^{ab}_\mu\,,\qquad
A^{a4}_\mu=\frac{1}{\ell}e^a_\mu\,,
\end{equation}
with the dimensionfull constant $\ell$ of dimension of length
introduced so as to keep the tetrad dimensionless for the canonical
dimension of connection. One then solves the algebraic equations for
the field $B^{IJ}$ and substitutes the result back to the action,
obtaining after some manipulations
\begin{eqnarray}
64 \pi \,
S&=&\frac{1}{G}\int\epsilon^{abcd}(R_{\mu\nu\,ab}e_{\rho\,c}e_{\sigma\,d}
-\frac{\Lambda}{3}e_{\mu\,a}e_{\nu\,b}e_{\rho\, c}e_{\sigma\, d})\epsilon^{\mu\nu\rho\sigma}\nonumber\\
&+&\frac{2}{G\gamma}\int R_{\mu\nu\,ab}\,e_{\nu}^{\,a}e_{\rho}^{\, b}\,\epsilon^{\mu\nu\rho\sigma}\label{6}\\
&+&\frac{\gamma^2+1}{\gamma\, G}NY_4
+\frac{3\gamma}{2G\Lambda}P_4-\frac{3}{4G\Lambda} E_4\, ,\nonumber
\end{eqnarray}
with
$$
\frac{\Lambda}{3}=-\frac{1}{\ell^2},\quad     \alpha =
\frac{G\Lambda}{3}\frac{1}{(1+\gamma^2)}, \quad \beta =
\frac{G\Lambda}{3}\frac{\gamma}{(1+\gamma^2)} , \quad
\gamma=\frac{\beta}{\alpha}.
$$
The first two terms in (\ref{6}) is just the Einstein-Cartan action
with the cosmological term, the third is the Holst term, and the
remaining ones are the Nieh-Yan, Pontryagin, and Euler invariants
(see eg., \cite{Freidel:2005ak} or \cite{Durka:2010zx} for details
of this construction.)

The action (\ref{6}) can be written down in a more compact form as
follows
\begin{equation}\label{4a}
 S(\omega,e)=\frac{1}{16\pi} \int\left( \frac{1}{4}M^{abcd} F_{ab}\wedge F_{cd}-\frac{1}{\beta\ell^2} \,T^a \wedge T_a\right)\,
 ,
\end{equation}
where the AdS curvature $F^{ab}$ is defined below (\ref{11}),
$T^a=F^{a4}$ is torsion, and
\begin{equation}\label{4b}
M^{ab}{}_{cd}=\frac{\alpha}{(\alpha^2+\beta^2)}( \gamma\,
\delta^{ab}_{cd}-\epsilon^{ab}_{\;\;cd}) \equiv -\frac{\ell^2}{G}(
\gamma\, \delta^{ab}_{cd}-\epsilon^{ab}_{\;\;cd})\,.
\end{equation}
We will find this particular form of the action convenient below.

Let us then repeat this construction in the case when the Anti de
Sitter symmetry is replaced with its Maxwell generalization. The
AdS-Maxwell algebra has the form (this algebra, which is a direct
sum of the Lorentz and anti de Sitter algebras,
$\sf{so}(3,1)\oplus\sf{so}(3,2)$, has been previously discussed in
\cite{Soroka:2006aj}, and \cite{Bonanos:2010fw}; see also
\cite{Lukierski:2010dy})
\begin{equation}\label{7}
[\mathcal{P}_{a},\mathcal{P}_{b}]=i(\mathcal{M}_{ab}-\mathcal{Z}_{ab})\,
,
\end{equation}
$$ [\mathcal{M}_{ab},\mathcal{M}_{cd}]=-i(\eta_{ac}\mathcal{M}_{bd}+\eta_{bd}\mathcal{M}_{ac}-\eta_{ad}\mathcal{M}_{bc}-\eta_{bc}\mathcal{M}_{ad}) ,$$
$$
[\mathcal{M}_{ab},\mathcal{Z}_{cd}]=-i(\eta_{ac}\mathcal{Z}_{bd}+\eta_{bd}\mathcal{Z}_{ac}-\eta_{ad}\mathcal{Z}_{bc}-\eta_{bc}\mathcal{Z}_{ad}),
$$
$$
[\mathcal{Z}_{ab},\mathcal{Z}_{cd}]=-i(\eta_{ac}\mathcal{Z}_{bd}+\eta_{bd}\mathcal{Z}_{ac}-\eta_{ad}\mathcal{Z}_{bc}-\eta_{bc}\mathcal{Z}_{ad}),$$
$$[\mathcal{M}_{ab},\mathcal{P}_c]=-i(\eta_{ac}\mathcal{P}_{b}-\eta_{bc}\mathcal{P}_{a}),\qquad [\mathcal{Z}_{ab},\mathcal{P}_{c}]=0\,.$$

One can readily gauge this algebra by defining the gauge field
(connection)
\begin{equation}\label{8}
\mathbb{A}_{\mu}=\frac{1}{2}\omega_\mu{}^{ab}
\mathcal{M}_{ab}+\frac{1}{\ell}e_\mu{}^{a}
\mathcal{P}_{a}+\frac{1}{2}h_\mu{}^{ab} \mathcal{Z}_{ab}
\end{equation}
and its curvature tensor
\begin{equation}\label{9}
\mathbb{F}_{\mu\nu}=\partial_\mu\mathbb{A}_{\nu}-\partial_\nu\mathbb{A}_{\mu}-i[\mathbb{A}_{\mu},\mathbb{A}_{\nu}]
\end{equation}
which can be decomposed into Lorentz, translational, and Maxwell
parts
\begin{equation}\label{10}
\mathbb{F}_{\mu\nu}=\frac12\, F_{\mu\nu}^{ab}\, \mathcal{M}_{ab}
+\frac{1}{\ell}T_{\mu\nu}^{a}\,\mathcal{P}_{a}+ \frac12\,
G_{\mu\nu}^{ab}\, \mathcal{Z}_{ab}\,,
\end{equation}
where
\begin{eqnarray}
F_{\mu\nu}^{ab}&=&R^{ab}_{\mu\nu}+\frac{1}{\ell^2}( e^a_\mu e^b_\nu- e^a_\nu e^b_\mu) \label{11}\\
T_{\mu\nu}^{a}&=&D^\omega_\mu e^a_\nu -D^\omega_\nu e^a_\mu\label{12}\\
G_{\mu\nu}^{ab}&=&D^\omega_\mu h^{ab}_\nu -D^\omega_\nu
h^{ab}_\mu-\frac{1}{\ell^2}( e^a_\mu e^b_\nu- e^a_\nu e^b_\mu)
+(h^{ac}_\mu h^{\quad b}_{\nu\,c}-h^{ac}_\nu h^{\quad
b}_{\mu\,c})\label{13}\, .
\end{eqnarray}
In the formula above we denote by $D^\omega_\mu(\ast)\equiv
\partial_\mu(\ast)-i[\omega_\mu,(\ast)]$ the covariant
derivative of the Lorentz connection $\omega$. Using the full
covariant derivative $D_\lambda^{\mathbb A}$ we can write the
Bianchi identity for the curvature $\mathbb{F}_{\mu\nu}$, to wit
\begin{equation}\label{14}
 \epsilon^{\mu\nu\rho\sigma}   D_\mu^{\mathbb A} \mathbb
 F(\mathbb{A})_{\nu\rho}=0\, ,
\end{equation}
which can be again decomposed into
\begin{eqnarray}
\epsilon^{\mu\nu\rho\sigma} \,D_\mu^{\omega}R_{\nu\rho}^{ab}=0\label{15}\\
\epsilon^{\mu\nu\rho\sigma} \,(D_\mu^{\omega}T_{\nu\rho}^a-R_{\mu\nu}^{ab} e_{\rho b})=0\label{16}\\
\epsilon^{\mu\nu\rho\sigma}\,(D_\mu^{(\omega+h)}G_{\nu\rho}^{ab}+2h_\mu^{ac}
F_{\nu\rho c}{}^{b}-\frac{2}{\ell^2} e_\mu^a
T_{\mu\nu}^b)=0\label{17}
\end{eqnarray}
Let us notice in passing that using (\ref{15})  the last identity
can be rewritten in a more compact form as follows
\begin{equation}\label{18}
\epsilon^{\mu\nu\rho\sigma}\,
D_\mu^{(\omega+h)}(G_{\nu\rho}^{ab}+F_{\nu\rho}^{ab})=\epsilon^{\mu\nu\rho\sigma}\,D_\mu^{(\omega+h)}R_{\nu\rho}^{ab}(\omega+h)=0
\end{equation}

Before turning to the construction of the action we need an explicit
form of gauge transformations of the components of connection and
curvature. These gauge transformation read
\begin{equation}\label{19}
\delta_\Theta \mathbb{A}_{\mu}=\partial_\mu \Theta
-i[\mathbb{A}_{\mu}, \Theta]\equiv D^{\mathbb{A}}_\mu \Theta\, ,
\qquad \delta_\Theta
\mathbb{F}_{\mu\nu}=i[\Theta,\mathbb{F}_{\mu\nu}]\, ,
\end{equation}
where the gauge parameter $\Theta$ decomposes into parameters of
local Lorentz, translation and Maxwell symmetries
\begin{equation}\label{20}
    \Theta=\frac{1}{2}\lambda^{ab}\mathcal{M}_{ab}+\xi^a \mathcal{P}_a+\frac{1}{2}\tau^{ab}\mathcal{Z}_{ab}\, .
\end{equation}
By direct calculation we see that the connection components
transform as follows
\begin{eqnarray}
        \delta_\Theta h^{ab}_{\mu}&=&D^\omega_\mu \tau^{ab}-\frac{1}{\ell}(e^a_\mu\,\xi^b-e^b_\mu\, \xi^a)
        +
h^{ac}_\mu(\lambda_{c}^{\;\,b}+\tau_{c}^{\;\,b})+h^{bc}_\mu(\lambda^{a}_{\;\;c}+\tau^{a}_{\;\;c})\nonumber\\
      \delta_\Theta \omega^{ab}_{\mu}&=& D^\omega_\mu \lambda^{ab}+\frac{1}{\ell}(e^a_\mu\, \xi^b-e^b_\mu\, \xi^a)
      \label{21}\\
    \frac{1}{\ell}\delta_\Theta e^{a}_{\mu}&=&D^\omega_\mu \xi^a-\frac{1}{\ell}\lambda^a_{\;b} \,e^b_\mu\,,\nonumber
\end{eqnarray}
while for the components of the curvature we find
\begin{eqnarray}
        \delta_\Theta G^{ab}_{\mu\nu}&=&\frac{1}{\ell}[\xi,T_{\mu\nu}]^{ab}-[\tau,F_{\mu\nu}]^{ab}-[(\lambda+\tau),G_{\mu\nu}]^{ab}\nonumber\\
      \delta_\Theta F^{ab}_{\mu\nu}&=&-\frac{1}{\ell}[\xi,T_{\mu\nu}]^{ab}-[\lambda,F_{\mu\nu}]^{ab}  \label{22}\\
    \frac{1}{\ell}\delta_\Theta T^{a}_{\mu\nu}&=&-\frac{1}{\ell}\lambda^a_{\;b}T^b_{\mu\nu}+\xi_b\,F^{ab}_{\mu\nu}\,.\nonumber
\end{eqnarray}
Let us now turn to the construction of the AdS-Maxwell analogue of
the action (\ref{4}). The generalization of the first term in
(\ref{4}) should look like $2B_a\wedge T^a + B_{ab}\wedge F^{ab} +
C_{ab}\wedge G^{ab}$, with $B^a=B^{a4}$.  These combination of terms
must be invariant under action of all local symmetries of the
theory, and this requirement fixes the transformation rules for the
fields $B$ and $C$ to be as follows
\begin{equation}\label{23}
 \delta_\xi  B^{ab}=({B}^{a} \xi^b-{B}^{b} \xi^a)\,, \quad \delta_\xi C^{ab}=0\,,\quad \delta_\xi
 {B}^{a}=(B^{ab}-C^{ab})\xi_b\,;
\end{equation}
\begin{equation}\label{24}
    \delta_\lambda B^{ab}=-[\lambda,B]^{ab}\,,\quad \delta_\lambda  C^{ab}=-[\lambda,C]^{ab}\,,\quad \delta_\lambda
    {B}^{a}=-\lambda^a{}_{b}{B}^{b}\,;
\end{equation}
\begin{equation}\label{25}
    \delta_\tau B^{ab}=-[\tau,C]^{ab} \,,\quad \delta_\tau C^{ab}=-[\tau,C]^{ab}\,,\quad  \delta_\tau
    {B}^{a}=0\,.
\end{equation}
In the next step we must generalize the second term in the action
(\ref{4}). Looking at (\ref{23})--(\ref{25}) we see that there are
two gauge invariant terms quadratic in the fields $B$ and $C$,
namely $$  B^a\wedge B_a +  B^{ab}\wedge B_{ab} - 2C^{ab}\wedge
B_{ab}\qquad \mbox{and}\qquad C^{ab}\wedge C_{ab}\,.$$

In the last step we must find the terms that are generalizations of
the third, gauge breaking term in (\ref{4}). Since here we are going
to differ from the choice of made in the paper
\cite{deAzcarraga:2010sw}, let us proceed with some care. In that
paper the authors allow for the cosmological constant term and all
terms linear and quadratic in curvatures, which were invariant under
local Lorentz transformations, not imposing any conditions following
from the Maxwell sector of the symmetry algebra. Here we follow a
different path, generalizing the last term in the action (\ref{4})
so as to preserve both the Lorentz and Maxwell symmetries.  In
another words we take the most general terms that break the
translational symmetry (which, as a result becomes the general
coordinate invariance on shell, as usual), keeping all the other
symmetries of the unconstrained theory operational. Since our
resulting action will have more symmetries than the one considered
in \cite{deAzcarraga:2010sw}, the dynamics it describes is expected
to be more restrictive than the one considered in that paper. As we
will see in a moment this is exactly what is going to happen. There
are two combinations of terms satisfying this requirement, namely
$\epsilon^{abcd}C_{ab}\wedge C_{cd}$ and
$\epsilon^{abcd}(B_{ab}\wedge B_{cd}-2C_{ab}\wedge B_{cd})$.
Therefore the action of our constrained topological theory has the
form
\begin{eqnarray}
 16\pi\, S(A,B)&=& \int 2(B^{a4}\wedge F_{a4}-\frac{\beta}{2}B^{a4}\wedge B_{a4}) \nonumber\\
& &  +B^{ab}\wedge F_{ab}- \frac{\beta}{2}B^{ab}\wedge B_{ab}- \frac{\alpha}{4}\epsilon^{abcd} B_{ab}\wedge B_{cd}\nonumber\\
& &+C^{ab}\wedge G_{ab}- \frac{\rho}{2}C^{ab}\wedge C_{ab} - \frac{\sigma}{4}\epsilon^{abcd} C_{ab}\wedge C_{cd}\nonumber\\
& & + \beta  C^{ab}\wedge B_{ab}+ \frac{\alpha }{2}\epsilon^{abcd}
C_{ab}\wedge B_{cd}\label{26}
\end{eqnarray}
By construction this action is invariant under local Lorentz and
Maxwell symmetries with the translational symmetry being broken
explicitly by the `epsilon' terms.

The algebraic $B$ and $C$ field equations take the form
\begin{eqnarray}
\frac1\ell\, T^{a}&=& \beta B^{a}\label{27}\\
G^{ab}&=&\rho C^{ab}+ \frac{\sigma}{2}\epsilon^{abcd} C_{cd}-\beta B^{ab}-\frac{\alpha}{2}\epsilon^{abcd} B_{cd}\label{28}\\
F^{ab}&=& \beta B^{ab}+ \frac{\alpha}{2}\epsilon^{abcd} B_{cd}-\beta
C^{ab}-\frac{\alpha}{2}\epsilon^{abcd} C_{cd}\label{29}
\end{eqnarray}
Using these equations the action (\ref{26}) can be written in the
simpler form
\begin{equation}\label{30}
 16\pi \, S(A,B)= \frac{1}{2}\int \left(B^{ab}\wedge F_{ab} + C^{ab}\wedge G_{ab}+\frac{2}{\beta}\, B^{a4}\wedge
 F_{a4}\right)\,,
\end{equation}
which after substituting the algebraic equations for $B$ and $C$
fields becomes
\begin{eqnarray}
 16 \pi&S&(\omega,h,e)= \int\left( \frac{1}{4}M^{abcd} F_{ab}\wedge F_{cd}-\frac{1}{\beta\ell^2} \,T^a \wedge T_a\right)\nonumber\\
 & &\quad+\int\frac{1}{4}N^{abcd} (F_{ab}+G_{ab})\wedge
 (F_{cd}+G_{cd})\label{31}
\end{eqnarray}
with $M^{abcd}$ given by (\ref{4b}) and
\begin{equation}\label{32}
    N^{abcd}=\frac{(\sigma-\alpha)}{(\sigma-\alpha)^2+(\rho-\beta)^2}\left(\frac{\rho-\beta}{\sigma-\alpha}\delta^{abcd} -\epsilon^{abcd} \right)\,.
\end{equation}
The action (\ref{31}) is the final result of this paper. Let us turn
to the discussion of its meaning.

The first line of (\ref{31}) is just our original action for gravity
with negative cosmological constant (and with Holst and topological
terms) given by eq.\ (\ref{4a}) and (\ref{6}). It is easy to see
that the second line of this expression is just a topological
invariant, which, in particular, does not contribute to the
dynamical field equations. This follows from the fact that the sum
of two curvatures $F^{ab}$ and $G^{ab}$ is the Riemannian curvature
of the sum of two connections
$$
F^{ab}(\omega,e)+G^{ab}(h,e) = R^{ab}(\omega+h)\equiv
d(\omega+h)^{ab} + (\omega+h)^{a}{}_c\wedge(\omega+h)^{cb}\, ,
$$
and, in particular the tetrad terms cancel out in this expression.
Therefore the term in the second line line of (\ref{31}) is a sum of
the Euler and Pontryagin invariants, calculated for the connection
$\omega+h$.

Thus we see that our construction leads just to the Einstein-Cartan
gravity action with the gauge field associated with the Maxwell
symmetry not influencing the dynamics and contributing only to the
boundary terms. In particular the Maxwell terms do not contribute to
the cosmological constant term and we do not see any trace of the
``generalized cosmological term'' described in
\cite{deAzcarraga:2010sw}.

However the disappearance of the field $h$ from the dynamics of the
theory is puzzling and requires explanation. Indeed, the $\tau$
gauge invariance in (\ref{21}) is not sufficient to gauge away the
field $h$. The resolution of this puzzle is simple. Our starting
constrained BFCG theory is geometrical. Its building blocks are one
forms and the only operations available in the construction of the
action are differentiation  $d$ and the wedge product $\wedge$ of
forms. Using these one cannot construct Yang-Mills terms in the
action, which require the use of the Hodge dual. Thus with the tools
at hands one simply cannot construct terms in the action that would
result in a nontrivial dynamics of the Maxwell field $h_\mu{}^{ab}$.

One could add such terms to the action (\ref{31}) by hands, of
course. It is easy to check that the lowest order dynamical term for
$h_\mu{}^{ab}$ that preserves both local Lorentz and Maxwell
symmetries would be of the form \cite{Soroka:2011tc}
\begin{equation}\label{33}
    e\, \big( F_{\mu\nu}{}^{ab} + G_{\mu\nu}{}^{ab}\big) \big( F^{\mu\nu}{}_{ab} +
    G^{\mu\nu}{}_{ab}\big)\, ,
\end{equation}
which, contrary to the terms in our geometrical action above is
non-polynomial in fields and would lead to the higher derivative
theory of gravity \cite{Stelle:1977ry}. It is tempting to speculate
that perhaps adding the Maxwell-gravity terms like (\ref{33}) may
render the behavior of a quantum theory defined by (\ref{31}),
(\ref{33}) less pathological \cite{Stelle:1976gc}. The theory
defined by the sum of actions (\ref{31}) and (\ref{33}) seems to be
quite interesting and we will discuss it in details in a separate
paper.

\section*{ACKNOWLEDGEMENTS}
We thank  J.\ Lukierski, D.V.\ Soroka, and V.A.\ Soroka for comments
and bringing some references to our attention.

The work of J.\ Kowalski-Glikman was supported in part by the grant
182/N-QGG/2008/0,  the work of R.\ Durka was supported by the
National Science Centre grant  N202 112740, and the work of R.\
Durka and M.\ Szczachor was supported by European Human Capital
Programme.


\begin{thebibliography}{99.}
\bibitem{Bacry:1970ye}
  H.~Bacry, P.~Combe, J.~L.~Richard,
  ``Group-theoretical analysis of elementary particles in an external electromagnetic field. 1.
  the relativistic particle in a constant and uniform field,''
  Nuovo Cim.\  {\bf A67 } (1970)  267-299.

\bibitem{Schrader:1972zd}
  R.~Schrader,
  ``The maxwell group and the quantum theory of particles in classical homogeneous electromagnetic fields,''
  Fortsch.\ Phys.\  {\bf 20 } (1972)  701-734.

\bibitem{Galindo:1967}
  A.~Galindo,
  ``Lie algebra extensions of the Poincar\'e Algebra,''
  J.\ Math.\ Phys.\  {\bf 8 } (1967)  768.

\bibitem{Soroka:2011tc}
  D.~V.~Soroka and V.~A.~Soroka,
  ``Gauge semi-simple extension of the Poincar\'e group,''
  arXiv:1101.1591 [hep-th].

\bibitem{Gomis:2009dm}
  J.~Gomis, K.~Kamimura and J.~Lukierski,
  ``Deformations of Maxwell algebra and their Dynamical Realizations,''
  JHEP {\bf 0908}, 039 (2009)
  [arXiv:0906.4464 [hep-th]].

\bibitem{Bonanos:2009wy}
  S.~Bonanos, J.~Gomis, K.~Kamimura, J.~Lukierski,
 ``Maxwell Superalgebra and Superparticle in Constant Gauge Backgrounds,''
  Phys.\ Rev.\ Lett.\  {\bf 104}, 090401 (2010).
  [arXiv:0911.5072 [hep-th]].

\bibitem{deAzcarraga:2010sw}
  J.~A.~de Azcarraga, K.~Kamimura, J.~Lukierski,
  ``Generalized cosmological term from Maxwell symmetries,''
  Phys.\ Rev.\  {\bf D83}, 124036 (2011).
  [arXiv:1012.4402 [hep-th]].


\bibitem{Starodubtsev:2003xq}
  A.~Starodubtsev,
  ``Topological excitations around the vacuum of quantum gravity. I: The
  symmetries of the vacuum,''
  arXiv:hep-th/0306135.

\bibitem{Smolin:2003qu}
  L.~Smolin and A.~Starodubtsev,
  ``General relativity with a topological phase: An action principle,''
  arXiv:hep-th/0311163.

\bibitem{Freidel:2005ak}
  L.~Freidel and A.~Starodubtsev,
  ``Quantum gravity in terms of topological observables,''
  arXiv:hep-th/0501191.

\bibitem{Wise:2006sm}
  D.~K.~Wise,
  ``MacDowell-Mansouri gravity and Cartan geometry,''
  arXiv:gr-qc/0611154.

\bibitem{Durka:2010zx}
  R.~Durka and J.~Kowalski-Glikman,
  ``Hamiltonian analysis of SO(4,1) constrained BF theory,''
  Class.\ Quant.\ Grav.\  {\bf 27} (2010) 185008
  [arXiv:1003.2412 [gr-qc]].

\bibitem{Soroka:2006aj}
  D.~V.~Soroka, V.~A.~Soroka,
  ``Semi-simple extension of the (super)Poincare algebra,''
  Adv.\ High Energy Phys.\  {\bf 2009 } (2009)  234147.
  [hep-th/0605251].

\bibitem{Bonanos:2010fw}
  S.~Bonanos, J.~Gomis, K.~Kamimura, J.~Lukierski,
  ``Deformations of Maxwell Superalgebras and Their Applications,''
  J.\ Math.\ Phys.\  {\bf 51 } (2010)  102301.
  [arXiv:1005.3714 [hep-th]].

\bibitem{Lukierski:2010dy}
  J.~Lukierski,
  ``Generalized Wigner-Inonu Contractions and Maxwell (Super)Algebras,''
  Proc.\ Steklov Inst.\ Math.\  {\bf 272} (2011) 1
  [arXiv:1007.3405 [hep-th]].

\bibitem{Stelle:1977ry}
  K.~S.~Stelle,
  ``Classical Gravity with Higher Derivatives,''
  Gen.\ Rel.\ Grav.\  {\bf 9 } (1978)  353-371.

\bibitem{Stelle:1976gc}
  K.~S.~Stelle,
  ``Renormalization of Higher Derivative Quantum Gravity,''
  Phys.\ Rev.\  {\bf D16 } (1977) 953-969.



\end{thebibliography}
\end{document}